\documentclass[final]{siamltex}
\usepackage{graphicx}
\usepackage{epsfig}
\title{An Efficient and User Privacy-Preserving Routing Protocol for Wireless Mesh Networks}

\author{Jaydip Sen\thanks{Innovation Lab, Tata Consultancy Services Ltd, Bengal Intelligent Park, Salt Lake Electronics Complex, Kolkata - 700091, INDIA. (jaydip.sen@tcs.com). Questions, comments, or corrections
to this document may be directed to that email address.}}

\begin{document}
\maketitle

\begin{abstract}
Wireless mesh networks (WMNs) have emerged as a key technology for next generation wireless broadband networks showing rapid progress and inspiring numerous compelling applications. A WMN comprises of a set of mesh routers (MRs) and mesh clients (MCs), where MRs are connected to the Internet backbone through the Internet gateways (IGWs). The MCs are wireless devices and communicate among themselves over possibly multi-hop paths with or without the involvement of MRs. User privacy and security have been primary concerns in WMNs due to their peer-to-peer network topology, shared wireless medium, stringent resource constraints, and highly dynamic environment. Moreover, to support real-time applications, WMNs must also be equipped with robust, reliable and efficient routing protocols so as to minimize the end-to-end latency. Design of a secure and efficient routing protocol for WMNs, therefore, is of paramount importance. In this paper, we propose an efficient and reliable routing protocol that also provides user anonymity in WMNs. The protocol is based on an accurate estimation of the available bandwidth in the wireless links and a robust estimation of the end-to-end delay in a routing path, and minimization of control message overhead. The user anonymity, authentication and data privacy is achieved by application of a novel protocol that is based on Rivest's ring signature scheme. Simulations carried out on the proposed protocol demonstrate that it is more efficient than some of the existing routing protocols. 
\end{abstract}


\begin{AMS}\end{AMS}

\pagestyle{myheadings}
\thispagestyle{plain}
\markboth{JAYDIP SEN}{An Efficient and User Privacy-Preserving Routing Protocol for Wireless Mesh Networks}

\section{Introduction}\label{Section1}

Wireless mesh networking has emerged as a promising concept to meet the challenges in next-generation wireless networks such as providing flexible, adaptive, and reconfigurable architecture while offering cost-effective solutions to service providers. WMNs are multi-hop wireless networks formed by mesh routers (which form a wireless mesh backbone) and mesh clients. The mesh routers provide a rich radio mesh connectivity which significantly reduces the up-front deployment cost of the network. Mesh routers are typically stationary and do not have power constraints. However, the clients are mobile and energy-constrained. Some mesh routers are designated as gateway routers which are connected to the Internet through a wired backbone. A gateway router provides access to conventional clients and interconnects ad hoc, sensor, cellular, and other networks to the Internet. A mesh network can provide multi-hop communication paths between wireless clients, thereby serving as a community network, or can provide multi-hop paths between the client and the gateway router, thereby providing broadband Internet access to the clients. 

As WMNs become an increasingly popular replacement technology for last-mile connectivity to the home networking, community and neighborhood networking, it is imperative to design an efficient resource management system for these networks. Routing is one of the most challenging issues in resource management for supporting real-time applications with stringent QoS requirements. However, most of the existing routing protocols for WMNs are extensions of protocols originally designed for \textit{mobile ad hoc networks} (MANETs) and thus they perform sub-optimally. 

This paper presents an efficient and secure routing protocol for WMNs that is able to handle stringent QoS requirements of real-time applications while providing user privacy in a secure way. It involves a very low control overhead and hence provides a high network throughput when the number of data sources in the network is large. While issues such as reduction of control overhead of routing and enhancement of network throughput have been addressed for WMNs in \cite{Ross1}, the protocol proposed in this paper is more efficient than those schemes as observed in the simulation results.

The key contributions of the paper are as follows: (i) It exploits network topological information to increase the efficiency of route discovery process and uses \textit{multi-point relay} (MPR) nodes and \textit{circular routing} (discussed in Section~\ref{Section4}) to enhance the network throughput by reducing the control overhead. (ii) It computes a reliable link quality estimator and utilizes it in route selection. (iii) It provides a framework for reliable and robust estimation of available bandwidth and end-to-end delay in a routing path so that flow admission with guaranteed QoS for applications can be made. It also ensures that the number of retransmission required is minimized. (iv) It provides a simple mechanism to identify selfish nodes who consume network resources but do not cooperate with other nodes in forwarding packets for others. (v) It presents a novel user anonymization scheme that enables secure authentication of the users while protecting their privacy.

The rest of this paper is organized as follows. Section~\ref{Section2} describes related work on routing in WMNs. Section~\ref{Section3} discusses some important challenges in routing in WMNs. Section~\ref{Section4} describes the details of the proposed routing protocol. Simulation results are presented in Section~\ref{Section5}. Finally, Section~\ref{Section6} concludes the paper while highlighting some future scope of work.  
  
\section{Related Work}\label{Section2}
Although significant amount of work has been done on routing in MANETs, very little work has been done for WMNs. Most of the routing protocols for MANETs such as AODV and DSR use hop-count as the routing metric. However, this is approach not well-suited for WMNs. The basic idea in minimizing the hop-count is that it reduces delay and maximizes the throughput. But the assumption here is that the links in the path are either perfect or do not work at all, and all links are of equal bandwidth. A routing scheme that uses the hop-count metric does not take link quality into consideration. A minimum hop-count path has, on the average, longer links between the nodes present in the path compared to a higher hop-count path. This reduces the signal strength received by the nodes in that path and thereby increases the loss ratio at each link \cite{Couto}. Hence, it is always possible that a two-hop path with a good link quality provides higher throughput than a one-hop path with a poor link quality. Moreover, wireless links usually have asymmetric loss rate \cite{Aguqayo}. Hence, new routing metrics based on link quality are proposed such \textit{expected transmission count} (ETX),\textit{per-hop round-trip time} (RTT), and per-hop packet pair. 

Different approaches have been suggested by researchers for designing routing protocols for WMNs. In \cite{Badis}, a QoS routing over OLSR protocol has been proposed that takes into account metrics such as bandwidth and delay where the source node proactively changes a flow's next hop in response to the change in available bandwidth on its path. In \cite{Draves1}, the authors have proposed a \textit{link quality source routing} (LQSR) protocol. It is based on DSR and uses ETX as the routing metric. A new routing protocol called \textit{multi-radio link quality source routing} (MR-LQSR) is proposed in \cite{Draves2}. The process of neighbor node discovery and propagation of link metric are same as those in DSR. However, assignment of link weight and computation of the path weight is different. A QoS enabling routing algorithm for mesh-based wireless LAN architecture has been proposed in \cite{Xue1}, where the wireless users form an ad hoc peer-to-peer network. The authors also have proposed a protocol for MANET called \textit{ad hoc QoS on-demand routing} (AQOR) \cite{Xue2}. In \cite{Yang1}, the authors have shown that if a \textit{weighted cumulative expected transmission time} \cite{Draves1} is used in a link state routing protocol, it does not satisfy the \textit{isotonicity} property of the routing protocol and leads to formation of routing loops. To avoid routing loops, an algorithm is proposed that uses \textit{metric of interference and channel switching} (MIC) as the routing metric. The \textit{MeshCluster} architecture \cite{Ramachandran} addresses important issues in WMNs such as auto-configuration of mesh and client nodes, routing and load balancing in the infrastructure. The routing is performed via AODV-ST, a protocol that proactively maintains spanning trees rooted at the gateways. The mobility of the clients is managed by DHCP protocol.

Some routing protocols for WMNs have been developed by extending the existing routing protocols for MANETS with gateway discovery functionality \cite{Chen}\cite{Hwang}\cite{Ahlund}\cite{Ratnachandani}\cite{Sun} \cite{Michalak} \cite{Baumann1}. Since these protocols provide unicast routes, individual routes must be maintained between every mobile node and one of the gateways. Therefore, these protocols scale poorly to the number of nodes in the mesh network.

In \cite{Cheng}, scalability to the number of mesh nodes is improved with the uses of location information. However, this kind of information is typically not available in scenarios where the mobile nodes in the mesh network are commodity laptops or hand-held devices.

A problem that is common to most of the routing protocols for MANETs and WMNs is that gateway announcements or gateway requests are prone to vanish due to route breaks, and the recovery procedure is often as expensive as establishing a new route. In contrast, \cite{Mosko1} proposes an efficient mechanism to fix broken routes locally.  Mosko et al. \cite{Mosko2} propose to establish multiple non-disjoint paths for better performance, but again the established routes are unicast and this protocol is not scalable to the number of mesh nodes. 

IN \cite{Ju}, a single-hop mesh network architecture has been proposed where mobile clients connect directly to the gateways. However, this approach requires a much higher mesh node density for a comparable wireless coverage. In \cite{Baumann2}, the authors have proposed an anycast routing (i.e. routing form any mobile node to any gateway in the network) protocol that is designed to scale to the network size and to be robust to node mobility.

In contrast to the above approaches, the proposed protocol performs an on-demand route discovery using multiple metrics like bandwidth, delay, and reliability of the links and provides a routing framework that can support high network throughput with a minimum control overhead.

\section{Routing Challenges in WMNs}\label{Section3}
This section first presents the generic architecture of a WMN and then discusses some specific challenges in designing routing algorithms for such networks.

\begin{figure}[htb]
%\vspace{2.5in}
\centering
\includegraphics[width=25pc]{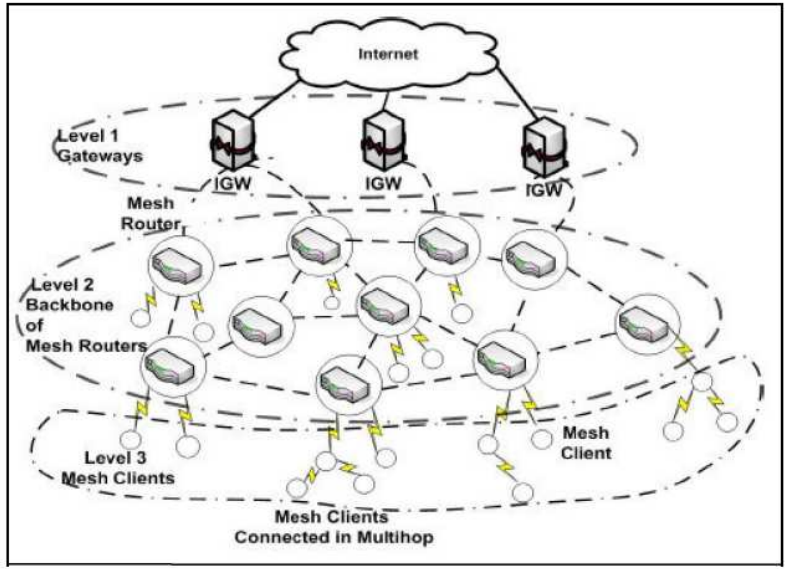}
\caption {The three-tier architecture of a wireless mesh network (WMN)}
\label{Figure1}
\end{figure} 

The architecture of a hierarchical WMN consists of three layers as shown in Fig.~\ref{Figure1}. At the top layers are the \textit{Internet gateways} (IGWs) that are connected to the wired Internet. They form the backbone infrastructure for providing Internet connectivity to the elements in the second level. The entities at the second level are called wireless \textit{mesh routers} (MRs) that eliminate the need for wired infrastructure at every MR and forward their traffic in a multi-hop fashion towards the IGW. At the lowest level are the \textit{mesh clients} (MCs) which are the wireless devices of the users. Internet connectivity and peer-to-peer communications inside the mesh are two important applications for a WMN. Therefore design of an efficient and low-overhead routing protocol that avoids unreliable routes, and accurately estimate the end-to-end delay of a flow along the path from the source to the destination is a major challenge.

(i) \textit{Measuring link reliability}: It has been observed that in wireless ad hoc networks nodes receiving broadcast messages introduce \textit{communication gray zones} \cite{Lundgren}. In such zones, data messages cannot be exchanged although the \textit{hello} messages reach the neighbors. This leads to a disruption in communication among the nodes. Since the routing protocols such as AODV and WMR \cite{Xue1} rely on control packets like RREQ, these protocols are highly unreliable for estimating the quality of wireless links. Due to communication gray zone problem, nodes that are able to send and receive bidirectional RREQ packets sometimes cannot send/receive data packets at high rate. These fragile links trigger link repairs resulting in high control overhead.

(ii) \textit{End-to-end delay estimation}: An important issue in a routing protocol is end-to-end delay estimation. Current protocols estimate end-to-end delay by measuring the time taken to route RREQ and RREP packets along the given path. However, RREQ and RREP are different from normal data packets and hence they are unlikely to experience the same levels of delay and loss as data packets. It has been observed through simulation that a RREP-based estimator overestimates while a hop-count-based estimator underestimates the actual delay experienced by the data packets \cite{Kone}. The reason for the significant deviation of a RREP-based estimator from the actual end-to-end delay is interference of signals. The RREQ packets are flooded in the network resulting in a heavy burst of traffic. This heavy traffic causes inter-flow interference in the paths. The unicast data packets do not cause such events. Moreover, as a stream of packets traverse along a route, due to the broadcast nature of wireless links, different packets in the same flow interfere with each other resulting in per-packet delays. Since the control packets do not experience per-packet delay, the estimate based on control packet delay deviate widely from the actual delay experienced by the data packets.

(iii) \textit{Reduction of control overhead}: Since the effective bandwidth of wireless channels vary continuously, reduction of control overhead is important in order to maximize throughput in the network. Reactive protocols like AODV and DSR use flooding of RREQ packets for route discovery. This consumes a high proportion of the network bandwidth and reduces the effective throughput. An important challenge in designing a routing protocol for WMNs is to optimize the communication and computation overhead of the control messages so that the bandwidth of the wireless channels may be used for applications as efficiently as possible. Security and privacy issues bring another dimension of complexity. The goal of the protocol designer would be to design the security framework in such a way that it involves minimum computational and message overhead.

\section{The Proposed Routing Protocol}\label{Section4}     
The goal of the proposed routing protocol is to establish a route from a source node to a destination node that allows traffic flow within a guaranteed end-to-end latency and with a guaranteed available bandwidth in the wireless channel using the minimum control overhead. The salient features of the proposed algorithm in this paper are now discussed in the following subsections. 

\subsection{Estimating Reliability of Routing Paths}\label{Section4.1}
Every node estimates the reliability of each of its wireless links to its one-hop neighbor nodes. For computing the reliability of a link, the number of control packets that a node receives in a given time window is used as a base parameter. An \textit{exponentially weighted moving average} (EWMA) method is used to update the link reliability estimate. If the percentage of control packets received by a node over a link in the last interval of measurement of link reliability is $N_t$, and if $N_{t-1}$ is the historical value of the link reliability before the last measurement interval, $\alpha=0.5$ is the weighting parameter, then the updated link reliability ($R$) is computed as in (\ref{equation1}):

\begin{equation}
R = \alpha . N_t + (1 - \alpha) . N_{t-1}\label{equation1}
\end{equation}

Every node maintains estimates the reliability of each of its links with its neighbors in a \textit{link reliability table}. The reliability for an end-to-end routing path is computed by taking the average of the reliability values of all the links on the path. The use of path with the highest reliability reduces the overhead of route repair. The paths with reliability values less than 0.5 are never selected for routing.

\subsection{Use of Network Topological Information in Route Discovery}\label{Section4.2}
The proposed protocol makes use of the knowledge of network topology by utilizing selective flooding of control messages in a portion of the network.  In this way, broadcasting of control messages is avoided and thus the chances of network congestion and disruption to the flows in the network are reduced. If both the source and the destination are under the control of the same mesh router (refer Fig.~\ref{Figure1}), the flooding of the control messages are confined within the portion of the network served by the mesh router only. However, if the source and the destination are under different mesh routers, the control traffic is limited to the two mesh groups.

To further reduce the overhead of control message and enhance the reliability in routing, the nodes accept broadcast control messages from only those neighbors which have the link reliability value greater than 0.5 (i.e., on average 50\% of the control packets sent from those nodes have been received by the node). This ensures that path with less reliability values are not discovered and therefore not considered for routing.

\subsection{Estimation of End-to-End Delay in a Routing Path}\label{Section4.3}
For accurate estimation of end-to-end delay in a routing path, an approach similar to the one proposed in \cite{Kone} has been taken. For addressing the issue of differential delays experienced by the control and the data packets, the proposed protocol makes use of some \textit{probe packets} during the route discovery phase. When a source node receives RREP packets from the destination in response to its RREQ, it stores in a table, the records for all the RREP packets together with the path through which the packets have arrived at it. However, instead of randomly selecting a path to send probe packets to the destination as suggested in \cite{Kone}, the packets are sent along the path from which the RREP messages have arrived at the source first. This ensures that the probe packets are sent along the path which is likely to induce less end-to-end delay resulting in a better performance of the protocol as observed from the simulation results presented in Section~\ref{Section5}. The probe packets are identical to data packets so far as their size, priority and flow rates are concerned. The objective of sending probe packets is to simulate the data flow and observe the delay characteristics in the routing path.  Number of probe packets is kept limited to $2H$ for a path consisting of $H$ hops to make a tradeoff between control overhead and measurement accuracy.

A destination node sets on a timer after it receives the first probe packet from the source node. The timer duration is based on the estimated time for receiving all the probe packets and is computed statistically. The destination computes the average delay experienced by all the probe packets it has received, and sends the computed value to the source node piggybacking it on a RREP message. If the computed value is within the limit of tolerance of the application QoS, the source selects the route and sends data packets. If the delay exceeds the required limit, the source selects the next best path (based on the arrival of RREP packets) from its table and tries once again. Since the routing path is set up based on the probe packets rather than the naïve RREP packets, the proposed protocol has higher route establishment The proposed algorithm has higher setup time due to sending of the probe packets and selection of the best path based on the estimated end-to-end delay. However, since the selected paths have high end-to-end reliability, the delay and the control packet overhead are reduced because of minimal subsequent route breaks.

\subsection{Use of Multi-Point Relay Nodes}\label{Section4.4}
The proposed routing protocol uses the \textit{multi-point relay} (MPR) nodes like the \textit{optimized link state routing} (OLSR) protocol \cite{Clausen} in order to reduce the control overhead in routing. In order to under the concept of MPR let us consider Fig.~\ref{Figure1}. 

The control messages sent by an IGW, called GW\_INFO messages are never flooded throughout the entire WMN; they are transmitted inside the corresponding subnet (under a particular MR) only. A GW\_INFO message is processed by a node if and only if the neighbor which forwarded it has been validated as bi-directional (i.e., the sender is reachable by the receiver via the reverse link). The bi-directionality of a link is determined by appending the list of neighbors in the periodic \textit{hello} messages. In this way, if a node finds itself in the list of neighbors advertised by its own neighbor, the link is considered bi-directional.

This additional list of neighbors in the hello messages is used to compute the MPR of a node. The objective of identifying the MPRs is to minimize control packet overhead. When MPRs are used, it is not necessary to send a message to all the nodes in a network when that message is required to reach all the nodes. If we visualize the WMN as a connected graph, the objective is to find the minimum subset of nodes which covers the whole graph. With a denser network, the benefits of using MPRs are more prominent. The protocol presented in this paper exploits the advantages of MPRs in order to reduce the control overheads of RREQ messages.

\subsection{Estimating Available Network Bandwidth}\label{Section4.5}
In addition to computation of path reliability and use of MPRs, it is also necessary that the effective bandwidth in a routing path is reliably estimated. This is extremely important to support real-time applications since these applications require a guarantee for a minimum available bandwidth. In the proposed protocol, the available bandwidth in a wireless link is estimated using its end-to-end delay and loss of packets due to congestion. The packet-loss due to congestion in the link is estimated as follows.In addition to computation of path reliability and use of MPRs, it is also necessary that the effective bandwidth in a routing path is reliably estimated. This is extremely important to support real-time applications since these applications require a guarantee for a minimum available bandwidth. In the proposed protocol, the available bandwidth in a wireless link is estimated using its end-to-end delay and loss of packets due to congestion. The packet-loss due to congestion in the link is estimated as follows.

In a wireless link packet loss may happen due to two reasons: (i) loss due to faulty wireless links and (ii) loss due to network congestion. The \textit{radio link control} (RLC) layer segments an IP packet into several RLC frames before transmission, and reassembles them into an IP packet at the receiver side. An IP packet loss occurs when an RLC frame belonging to an IP packet fails to be delivered. When this happens, the receiver knows the RLC frames reassembly has failed and the IP packet has been lost due to wireless error. Meanwhile, the sender detects \textit{retransmission time out} (RTO) of the frame and discards all the RLC frames belonging to the IP packet. This enables the sender to compute packet drop rate in the wireless links. Moreover, using the sequence numbers of the IP packets received at the receiver, it possible to differentiate the packet loss due to link error and packet loss due to congestion \cite{Yang2}. For example, while receiving two incoming packets with sequence number $i$ and $i$ + 2, if the receiver finds an IP packet assembly failure in RLC layer, the packet with sequence number $i$ + 1 is lost due to wireless channel. Once the packet loss ratio due to congestion ($P_{congestion}$) is estimated, the available bandwidth in the wireless link, \textit{estrat}, is given by (\ref{equation2}) as computed in \cite{Yang2}:

\begin{equation}
estrat = \frac {Packet Size} {(X + Y)}\label{equation2}
\end{equation}

$X$ and $Y$ in (\ref{equation2}) are computed using (\ref{equation3}) and (\ref{equation4}) as follows:

\begin{equation}
X = RTT  \sqrt {\frac {2P_{congestion}}{3}}\label{equation3}
\end{equation}
 
\begin{equation}
Y = RTO * Min (1, 3 \sqrt{\frac{3P_{congestion}}{8}}P_{congestion}(1 + 32{P^2_{congestion}})\label{equation4}
\end{equation}

In (\ref{equation2}), RTT is the average round trip time for a control packet. RTO is the retransmission time out for a packet, and is computed using (\ref{equation5}).

\begin{equation}
RTO = \overline{RTT} + K.\overline{RTT_{Var}}\label{equation5}
\end{equation}

$\overline{RTT}$ and $\overline{RTT_{Var}}$ in are the mean and variance respectively of RTTs and $k$ is set to 4. This bandwidth estimator is employed to dynamically compute the available bandwidth in the wireless links on a routing path so that the guaranteed minimum bandwidth for the flow is always maintained throughout the application life-time.

\subsection{Routing through the Fixed Network}\label{Section4.6}
In the proposed algorithm, the routing efficiency is further enhanced by occasional routing of packets through the fixed wired network backbone. Since the wired network backbone provides higher available bandwidth with more reliable links, it is advantageously exploited for intra-mesh message communication. 

Since the IGWs (refer Fig.~\ref{Figure1}) periodically announce their presence in the network through beacon messages, every mesh client knows the hop count from itself to its selected gateway. In the proposed protocol, the RREQ messages include this hop count information. When the destination receives the RREQ, since it also knows its distance from its gateway, it checks whether it is better (in terms of number of hops) to route the packet through the wireless nodes (mesh) or through the fixed network. 

In the proposed protocol, if a destination node finds that the better route is through the fixed network, the RREP message is routed through the wired network using the default route. Therefore, in such situations, the forward route is established between the source and the destination through the wired network, while the reverse route is set up through the WMN. This approach is known as \textit{circular routing} \cite{Ross1}. This approach, improves the performance of bi-directional flows between a source and destination pair (as in a TCP connection) since the nodes in the forward and in the reverse routes are on node-disjoint paths and do not contend for access of the wireless medium.

\subsection{Identification of Selfish Nodes}\label{Section4.7}
The proposed routing protocol also enforces cooperation among the nodes by identifying the selfish nodes in the network and isolating them. Selfishness is an inherent problem associated with any capacity-constrained multi-hop wireless networks like WMNs. A mesh router can behave selfishly owing to various reasons such as: (i) to obtain more wireless or Internet throughput, or (ii) to avoid path congestion. A selfish mesh router increases the packet delivery latency, and also increases the packet loss rate. A selfish node while utilizing the network resources for routing its own packet, avoids forwarding packets for others to conserve its energy. Identification of selfish nodes is therefore, a vital issue. 

Several schemes proposed in the literature to mitigate the selfish behavior of nodes in wireless networks such as credit-based schemes, reputation-based schemes and game theory-based schemes \cite{Santhanam}. However, to keep the overhead of computation and communication at the minimum, the proposed protocol employs a simple mechanism to discourage selfish behavior and encourage cooperation among nodes. To punish the selfish nodes, each node forwards packets to its neighbor node for routing only if the link reliability of that node is greater than a threshold (set at 0.5). Since the link reliability of a selfish node is 0, the packets arriving from this node will not be forwarded. Therefore, to keep its link reliability higher than the threshold, each node has to participate and cooperate in routing. The link reliability serves a dual purpose of enhancing reliability and enforcing node cooperation in the network.

\subsection{User Anonymity and Privacy}\label{Section4.8}
As mentioned in Section~\ref{Section1}, the proposed protocol has been augmented with a security module that provides user anonymity and privacy. An \textit{authentication server} (AS) has been used in the network that authenticates the users in the WMN while preserving their privacy. To enable user authentication and anonymity, a novel protocol has been designed extending the improved ring signature authentication scheme in \cite{Cao}. 

It is assumed that a symmetric encryption algorithm $E$ exists such that for any key $k$, the function $E_{k}$ is a permutation over $b$-bit strings. We also assume the existence of a family of \textit {keyed combining functions} $C_{k,\nu}(y_{1}, y_{2},......,y_{n})$ \cite{Rivest}, and a publicly defined collision-resistant hash function $H(.)$ that maps arbitrary inputs to strings of constant length which are used as keys for $C_{k,\nu}(y_{1},y_{2},......., y_{n})$. Every keyed combining function $C_{k,\nu}(y_{1}, y_{2},......, y_{n})$ takes as input the key $k$, an initialization $b$-bit value $\nu$, and arbitrary values $y_{1}$, $y_{2}$......,$y_{n}$. A user $U_{i}$ who wants to generate a session key with the authentication server, uses a ring of $n$ logged-on-users and performs the following steps.

\textit{Step 1}: $U_{i}$ chooses the following parameters: (i) a large prime $p_{i}$ such that it is hard to compute the discrete logarithms in $GF(p_{i})$, (ii) another large prime $q_{i}$ such that $q_{i}$ / ($p_{i}$ -1), and (iii) a generator $g_{i}$ in $GF(p_{i})$ with order $q_{i}$.

\textit{Step 2}: $U_{i}$ chooses $x_{A}$ $\in$ $Z_{q_{i}}$ as his private key, and computes the public key $y_{A_{i}}$ = ${g_{i}}^{x_{A_{i}}}$ mod $p_{i}$.

\textit{Step 3}: $U_{i}$ defines a trap-door function: \[f_{i}(\alpha, \beta) = \alpha.y_{A_{i}}^{\alpha mod q_{i}}.g_{i}^{\beta}mod p_{i}\] Its inverse function $f_{i}^{-1}$($y$) is defined as: $f_{i}^{-1}$ = ($\alpha$, $\beta$), where $\alpha$ and $\beta$ are computed as in (\ref{equation6}), (\ref{equation7}), and (\ref{equation8}). In these equations, $K$ is a random integer $Z_{q_{i}}$).

\begin{equation}
\alpha = y_{A_{i}}.g_{i}^{-K.(g_{i}^{k}mod q_{i})}mod p_{i}\label{equation6}
\end{equation}
\begin{equation}
\alpha^* = \alpha mod q_{i}\label{equation7}
\end{equation}
\begin{equation}
\beta = K.(g_{i}^K mod p_{i}) - x_{A_{i}}.\alpha^{*} mod q_{i}\label{equation8}
\end{equation}

$U_{i}$ makes $p_{i}$, $q_{i}$, $g_{i}$ and $y_{A_{i}}$ public, and keeps $x_{A_{i}}$ as secret.

The \textit {authentication server} (AS) chooses: (i) a large prime $p$ such that it is hard to compute discrete logarithms in $GF(p)$, (ii) another large prime $q$ such that $q$ / ($p$ -1), (iii) a generator $g$ in $GF(p)$ with order $q$, (iv) a random integer $x_{B}$ from $Z_{q}$ as its private key. The \textit{AS} computes its public key $y_{B}$ = $g^{x_{B}}$mod $p$ and publishes ($y_{B}$,$p$,$q$,$g$).

\emph{Anonymous authenticated key exchange}: The key-exchange is initiated by the user $U_{i}$ and invloves three rounds to computes a secret session key between $U_{i}$ and \textit{AS}. The operations in these three rounds are as follows:

\textit{Round 1}: When $U_{i}$ wants to generate a session key on the behalf of $n$ ring users $U_{1}$, $U_{2}$,......$U_{n}$ where, ($1 \leq i \leq n$), $U_{i}$ does the following:

(i)\ $U_{i}$ chooses two random integers $x_{1}$, $x_{A}$ $\in$ $Z_{q}^{*}$ and computes the following: $R = g^{X_{1}}mod\ p$, $Q = y_{B}^{X_{1}}mod\ p\ mod\ q$, $X = g^{x_{A}}mod\ p$ and $l = H(X, Q, V, y_{B}, I)$.

(ii)\ $U_{i}$ chooses a pair of values ($\alpha_{t}, \beta_{t}$) for every other ring member $U_{t}$, ($1\leq t\leq n, t\neq k$) in a pseudorandom way, and computes $y_{t}=f_{t}(\alpha_{t},\beta_{t})mod\ p_{t}$.

(iii)\ $U_{i}$ randomly chooses a $b$-bit initialization value $\nu$, and finds the value of $y_{i}$ from the equation: $C_{k,\nu}(y_{1},y_{2},......,y_{n})=\nu$.

(iv)\ $U_{i}$ computes ($\alpha_{i},\beta_{i})=f_{i}^{-1}(y_{i})$ by using the trap-door information of $f_{i}$. First, it chooses a random integer $K \in Z_{q_{i}}$, computes $\alpha_{i}$ using , and keeps $K$ secret. It then computes $\alpha_{i}^{*}$ using (\ref{equation7}), and finally computes $\beta_{i}$ using (\ref{equation8}) .

(v)\ ($U_{1},U_{2},......U_{n},\nu,V, R, (\alpha_{1}, \beta_{1}),(\alpha_{2},\beta_{2})....,(\alpha_{n},\beta_{n})$) is the ring signature $\sigma$ on $X$.

\textit{Round 2}: \textit {AS} does the following to recover and verify $X$ from the signature $\sigma$.

(i)\ textit{AS} computes $Q = R^{X_{B}}mod\ p\ mod\ q$, recovers $X$ using $X = V.g^{Q}mod\ p$, and hashes $X$, $Q$, $V$ and $y_{b}$ to recover $l$, where $l= H(X,Q,V,y_{B},I)$.

(ii)\ textit{AS} computes $y_{t}=f_{i}(\alpha_{t},\beta{t})mod\ p_{i}$, for $t = 1,2,....,n$.

(iii)\ \textit{AS} checks whether $C_{k,v}(y_{1},y{2},......y_{n})=\nu$. If it is true, \textit{AS} accepts $X$ as valid; otherwise, \textit{AS} rejects $X$. If $X$ is valid, \textit{AS} chooses a random integer $x_{b}$ from $Z_{q}^{*}$, and computes the following:    If it is true, AS accepts X as valid; otherwise, AS rejects X. If X is valid, AS chooses a random integer xb from , and computes the following: $Y=g^{x_{b}}mod\ p$, $K_{s}=X^{x_{b}}mod\ p$, and $h=H(K_{s},X,Y,I)$. \textit{AS} sends \{$h,Y,I^{'}$\} to $U_{i}$. 

\textit{Round 3}: $U_{i}$ verifies whether $K_{S^{'}}$  is from the server \textit{AS}. For this purpose, $U_{i}$ computes $K_{S}^{'}=Y^{x_{a}}mod\ p$, hashes $K$, $X$, $Y$ to get $h_{'}$ using $h^{'}=H(K_{S}^{'},X,Y,I)$. If $h^{'}?=h$, $U_{i}$ accepts $K_{S}$ as the session key.

\textit{Security analysis}: The key exchange scheme satisfies the following requirements.

(i)\textit{User anonymity}: For a given signature $X$, the server can only be convinced that the ring signature is actually produced by at least one of the possible users. If the actual user does not reveal the seed $K$, the server cannot determine the identity of the user. The strength of the anonymity depends on the security of the pseudorandom number generator.  It is not possible to determine the identity of the actual user in a ring of size $n$ with a probability greater than $1/n$. Since the values of $k$ and $\nu$ are fixed in a ring signature, there are $(2^{b})^{n-1}$  number of ($x_{1},x_{2},......x_{n}$) that satisfy the equation $C_{k,\nu}(y_{1},y_{2},......y_{n})=\nu$, and the probability of generation of each ($x_{1},x_{2},......x_{n}$) is the same. Therefore, the signature can't leak the identity information of the user.

(ii)\textit{Mutual authentication}: In the proposed scheme, not only the server verifies the users, but the users can also verify the server. Because of the hardness of inverting the hash function $f(.)$, it is computationally infeasible for the attacker to determine $(\alpha_{i},\beta_{i})$ and hence it is infeasible for him to forge a signature. If the attacker wants to masquerade as the \textit{AS}, he needs to compute $h=H(K_{S},X,Y)$. He requires $x_{B}$ in order to compute $X$. However, $x_{B}$ is the private key of \textit{AS} to which the attacker has no access.  
(iii)\textit{Forward secrecy}: The forward secrecy of a scheme refers to its ability to defend leaking of its keys of previous sessions when an attacker is able to catch hold of the key of a particular session. The forward secrecy of a scheme enables it to prevent \textit{replay attacks}. In the proposed scheme, since $x_{a}$ and $x_{b}$ are both selected randomly, the session key of each period has not relation to the other periods. Therefore, if the session key generated in the period $j$ is leaked, the attacker can not get any information of the session keys generated before the period $j$. The proposed protocol is, therefore, resistant to replay attack.  

\section{Performance Evaluation}\label{Section5}
\begin{figure}[htb]
%\vspace{2.5in}
\centering
\includegraphics [width=25pc] {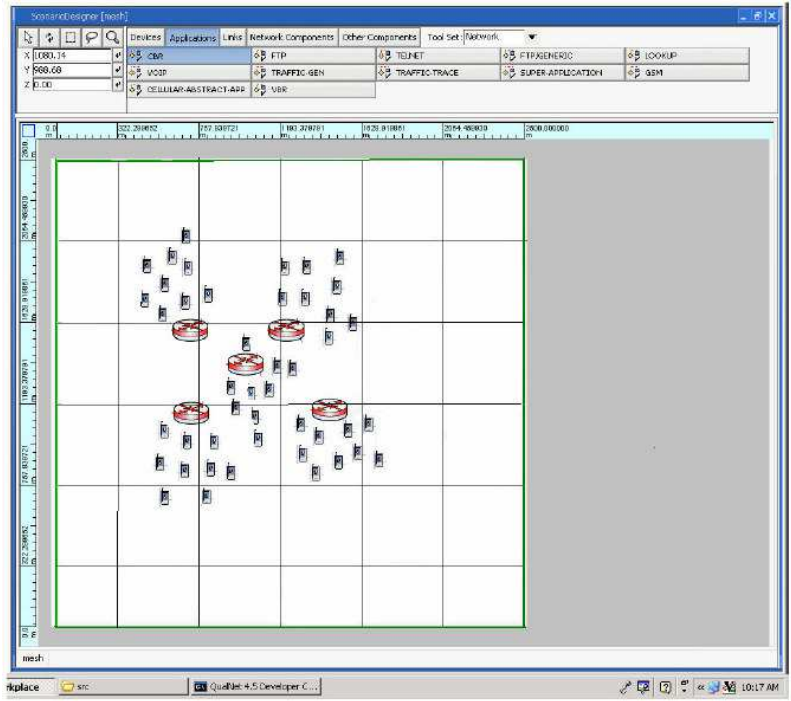}
\caption {The simulated WMN topology in Qualnet network simulator}
\label{Figure2}
\end{figure}

The proposed protocol has been implemented in the \textit{Qualnet} network simulator, version 4.5 \cite{Qualnet}. The reason for the choice of Qualnet is its ability of handle simulation of complex networks at multiple layers of the protocol stack.It is also suitable for simulation of large-scale dense networks like WMNs. The simulated network consists of 50 and 75 static nodes randomly distributed in the simulation area forming a dense WMN. The WMN topology is shown in Fig.~\ref{Figure2}  where 50 nodes are deployed in the network. out of these 50 nodes, 5 are MRs and remaining 45 are MCs. Each MR has 9 MCs associated with it. In 75 nodes deployment scenario (which is not shown), each of the 5 MRs has 14 MCs under it. 

For connectivity with the backbone Internet, 5 IGWs are placed at locations (100, 100), (100, 1400), (1400, 100) and (1400, 1400) and (700, 700) so as to provide uniform connectivity to backbone Internet with the WMN. The simulation parameters are presented in Table \ref{ParameterTable}. The choice of the parameters is made similar to that in \cite{Ross1}, and the performances of the two protocols (the proposed protocol and in one presented in \cite{Ross1}) are compared with respect to two important metrics- control packet overhead and network throughput.The overhead due to security and privacy module in the proposed protocol is not considered for the purpose of comparison, since unlike the protocol presented in this paper, the protocol in \cite{Ross1} does not have any security feature. The security and privacy module involves well-known symmetric key encryption, and computations of efficient hash functions and message digests.The computation and communication overhead invloved in these operations are well-known and hence they are not studied in the simulation. The simulation results are presented for the modules which are mainly responsible for efficient operation of the protocol, such as reduction of control message overhead, reduction of end-to-end latency and increasing the network throughput. 

\begin{table}[htb]
\caption{Simulation Parameters}
\label{ParameterTable} 
\begin{center} \footnotesize
\renewcommand{\tabcolsep}{2pc} % enlarge column spacing
\renewcommand{\arraystretch}{1.2} % enlarge line spacing
\begin{tabular}{@{}|l|l|} \hline 
\textbf{Parameter} &{\textbf{Values}}  \\ 
\hline
Simulated netwoork area		       	& 1500m * 1500m \\
Propagation channel frequency 		& 2.4 GHz \\
Raw channel bandwidth         		& 2 Mbps \\
MAC protocol								& 802.11b \\
Simulation duration           		& 900 s \\
Radio range of each node				& 250 m \\
Traffic type                  		& CBR UDP \\
Packet size                   		& 512 bytes \\
Data rate in the network      		& 32 Kbps   \\
IGW hello packet broadcast interval & 200 ms \\
No. of source nodes						& 15, 25, 35 \\
Node mobility								& None       \\
Wireless fading model					& None       \\
IP queue scheduler						& Strict priority \\
Propagation model							& Two-ray ground \\
Wired network bandwidth					& 100 Mbps \\
Delay in wired links						& 11.8 ms \\
\hline
\end{tabular}\\[2pt]
\end{center}
\end{table} 

\begin{figure}[htb]
%\vspace{2.5in}
\centering
\includegraphics [width=25pc] {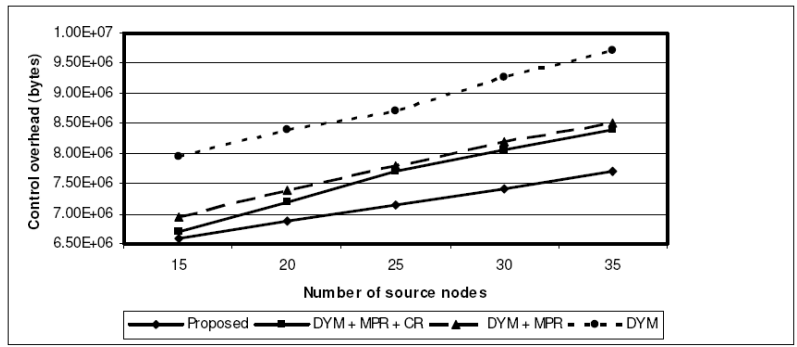}
\caption {Control overhead (bytes) vs. number of data source nodes (50 nodes in the networks)}
\label{Figure3}
\end{figure} 

\subsection{Control Overhead}\label{Section5.1}
\begin{figure}[htb]
%\vspace{2.5in}
\centering
\includegraphics [width=25pc]{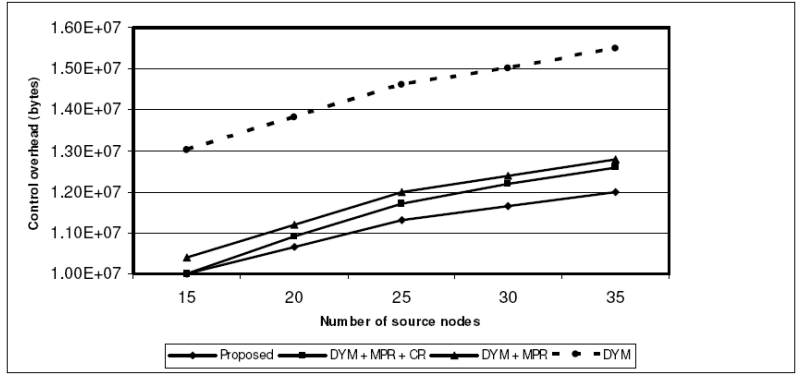}
\caption {Control overhead (bytes) vs. number of data source nodes (75 nodes in the networks)}
\label{Figure4}
\end{figure}

For studying the control overhead, four algorithms are considered. The DYNMOUM algorithm in \cite{Ross2}, the DYNMOUM with MPR, in \cite{Ross1} DYNMOUM with MPR and \textit{circular routing} (CR) \cite{Ross1} and the proposed algorithm compared with respect to their control overhead in routing.  The number of data source nodes is varied from 15 to 35 and the control overhead in bytes is studied for 50 nodes and 75 nodes networks respectively. The results are presented in Fig.~\ref{Figure3} and Fig.~\ref{Figure4}. 

It may be easily observed that the proposed protocol has the least control overhead among all the four protocols. The reason for the less control overhead in the proposed protocol is the less number of route errors and route repairs due the reliable link quality estimation and bandwidth estimation technique used in the protocol, which were absent in the other three protocols. In addition, it exploits the advantages of using MPRs and the circular routing. The MPRs reduce the overhead by controlled flooding and the circular routing reduces the overhead by routing some of the RREPs through the fixed network.

\begin{figure}[htb]
%\vspace{2.5in}
\centering
\includegraphics [width=25pc]{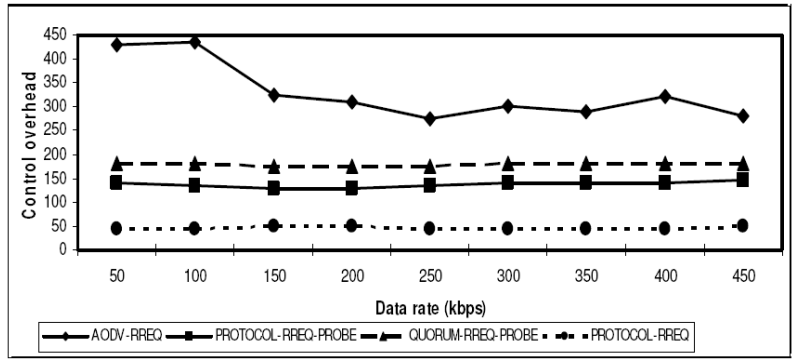}
\caption {Control overhead (bytes) vs. data rate (kbps). Comparison of performance of the proposed protocol with the QUORUM protocol in \cite{Kone}}
\label{Figure5}
\end{figure}  

To further demonstrate the efficiency of the proposed protocol, the control packets overhead of the protocol is also compared with that of the protocol presented in \cite{Kone}. For the purpose of comparison, some changes in the simulation parameters are made. The raw channel bandwidth is set at 11 Mbps. The application traffic is assumed to be CBR UDP. Each flow source is assumed to be sending a maximum of 10,000 packets to its destination node. Each flow is alive for 10 minutes and each simulation run is executed for 15 minutes. The link robustness value is computed once per second and the value of $\alpha$ in EWMA is taken as 0.5. All these parameters are set as per simulation environment presented in \cite{Kone}. For the purpose of comparison, only the RREQ messages and the probe packets in the protocols are considered since these broadcast messages largely contribute to the control overhead. Fig.~\ref{Figure5} shows the overhead due to RREQ packets in AODV and the proposed protocol for different data rates. The control overhead of the proposed protocol is first evaluated only with the RREQ packets and then with RREQ packets and the probe packets together. This also gives an idea of the additional overhead introduced due to the probe packets. It can be easily observed that the proposed protocol has very low overhead even with the probe packets when compared with the naive AODV protocol. It is also worth observing that the proposed protocol has about 20\% less control overhead that the protocol proposed in \cite{Kone} due to its robust bandwidth estimation of the wireless links.

\begin{figure}[htb]
%\vspace{2.5in}
\centering
\includegraphics [width=25pc]{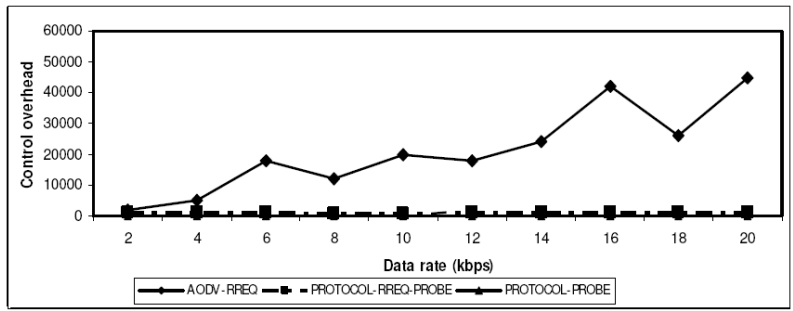}
\caption {Control overhead vs. number flows in the network. Proposed protocol has much less control overhead than AODV protocol and has similar performance as that of QUORUM protocol in \cite{Kone}}
\label{Figure6}
\end{figure} 

Fig.~\ref{Figure6} shows that AODV has a very high overhead due to control packets for increasing number of flows in the system. AODV always tries to establish routing paths based on minimum hop-counts. It does not consider the aspect of link reliability. This leads to frequent selection of unreliable links and consequent link-breaks and consequent re-discovery of routes resulting in high overhead of control packets. In contrast, the proposed protocol has a very limited control overhead since paths with higher link reliability only are selected for routing purpose. The algorithm proposed in \cite{Kone} has similar performance as the proposed protocol in this case. The results clearly demonstrate while the proposed protocol has similar trends as the protocol in \cite{Kone} as far as control overhead with number of flows in the network are concerned, it has almost 20\% reduction in control overhead for a particular value of network flow when compared with the protocol in \cite{Kone}. 

\begin{figure}[htb]
%\vspace{2.5in}
\centering
\includegraphics [width=25pc]{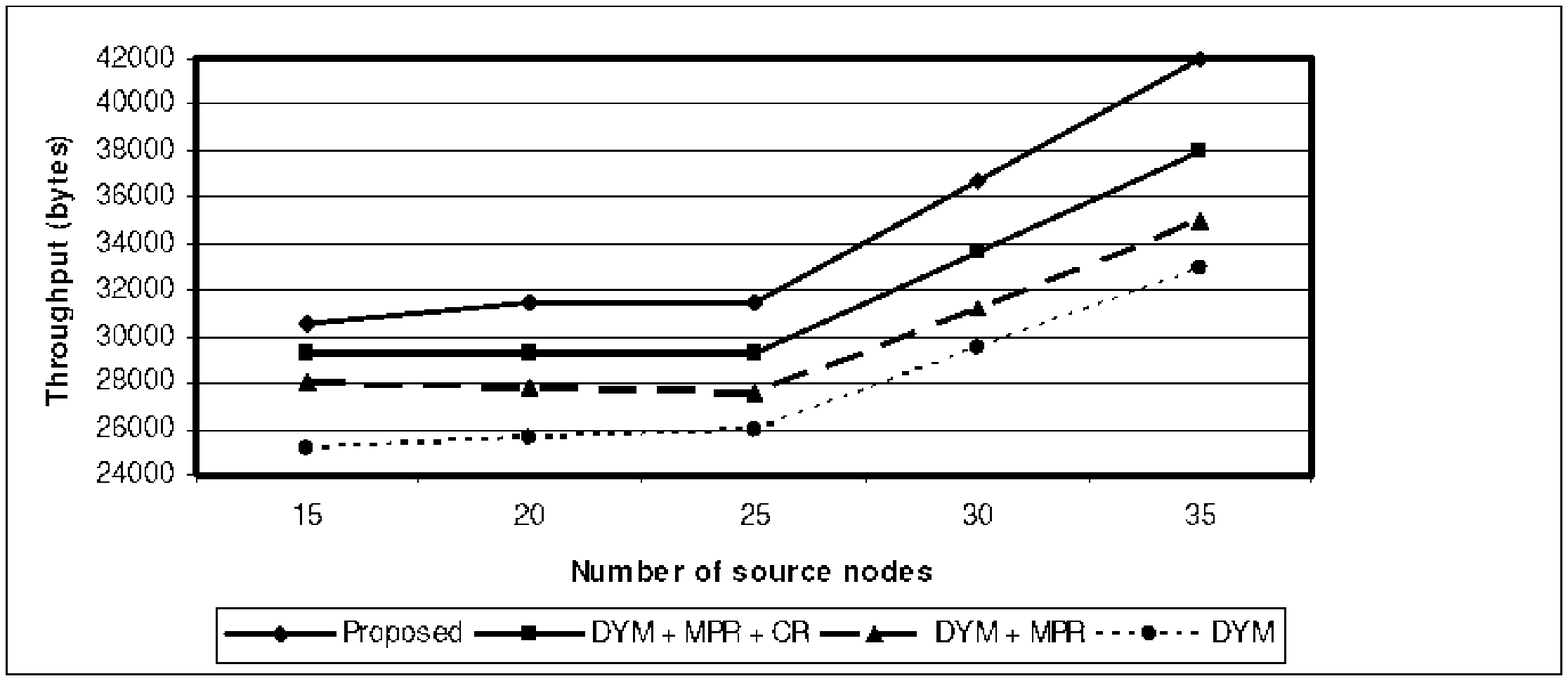}
\caption {Network throughput in bits per second (BPS) vs. number of data source nodes (50 nodes in the network)}
\label{Figure7}
\end{figure}  

\subsection{Network Throughput}\label{Section5.2}
\begin{figure}[htb]
%\vspace{2.5in}
\centering
\includegraphics [width=25pc]{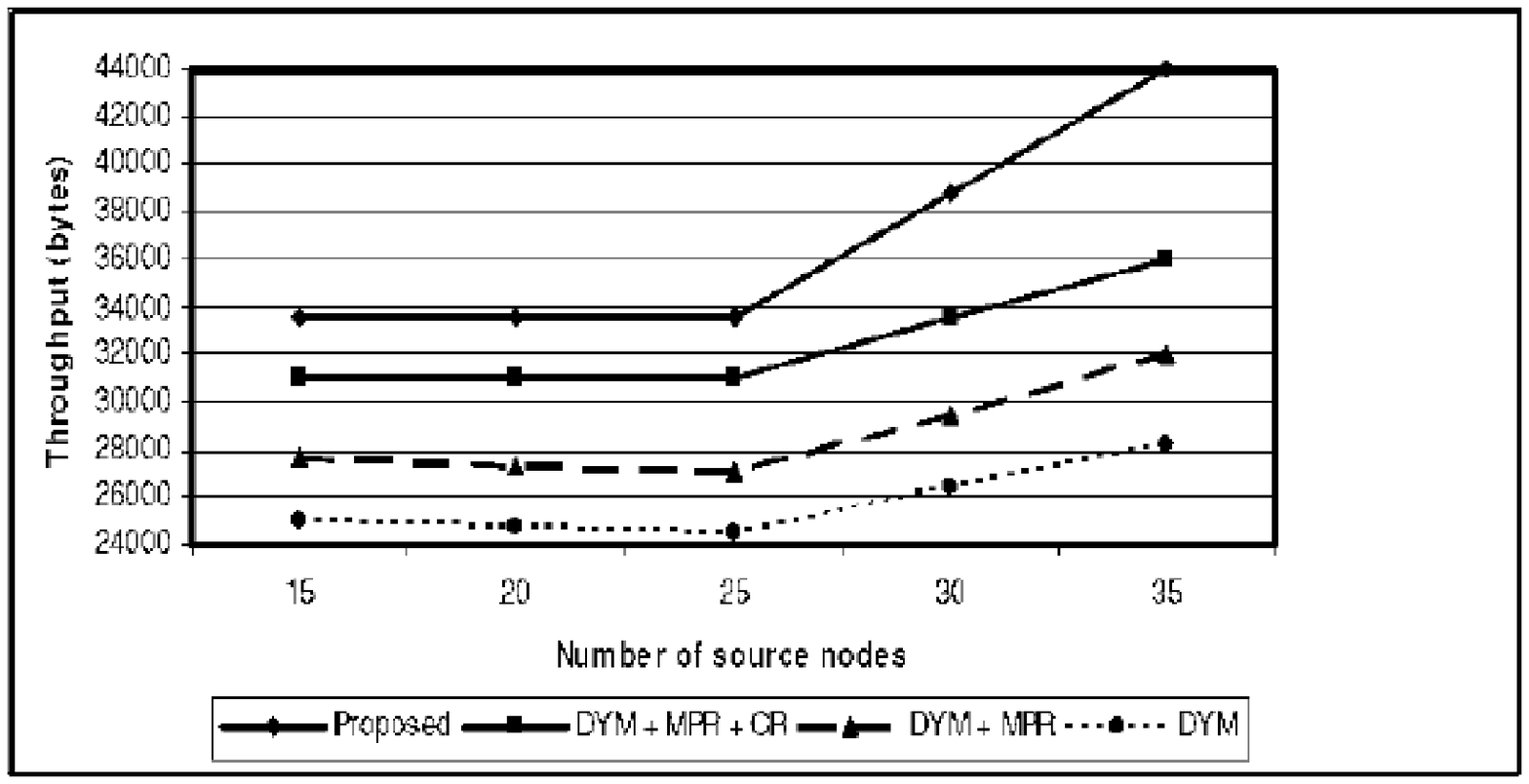}
\caption {Network throughput in bits per second (BPS) vs. number of data source nodes (75 nodes in the network)}
\label{Figure8}
\end{figure}

The performance of the protocol is also studied with respect to its ability to enhance network throughput. It may be intuitively clear that the reduction in control overhead should lead to a corresponding increase in the network throughput. Fig.~\ref{Figure7} and Fig.~\ref{Figure8} represent the data throughput in the network under varying number of source nodes with total number of nodes in the network being 50 and 75 respectively.

It may be observed that the proposed protocol produces maximum network throughput among all the four protocol studied. There are various factors that contribute to the enhanced throughput with the proposed protocol. First, the throughput significantly increases when MPRs are used due to reduced number of collision in the wireless medium and fewer retransmissions. Moreover, circular routing improves the performance further due to use of fixed network that has higher effective bandwidth. Accurate estimate of link quality also contributes to higher throughput, since packets are always forwarded through the link that has the highest effective bandwidth. Finally, efficient bandwidth estimation ensures that there will be minimum packet retransmission.

\subsection{End-to-End Delay Estimation}\label{Section5.3}
To demonstrate the effectiveness of the end-to-end delay estimation mechanism by probe packets mentioned in Section~\ref{Section4.3}, delays estimated by naïve RREP approach and the probe packet approach are compared with the actual end-to-end delay in the  routing paths. The records are observed for different flow rates with each flow having a minimum bandwidth requirement of $B_{min} = 50$ Kbps.

\begin{figure}[htb]
%\vspace{2.5in}
\centering
\includegraphics [width=25pc]{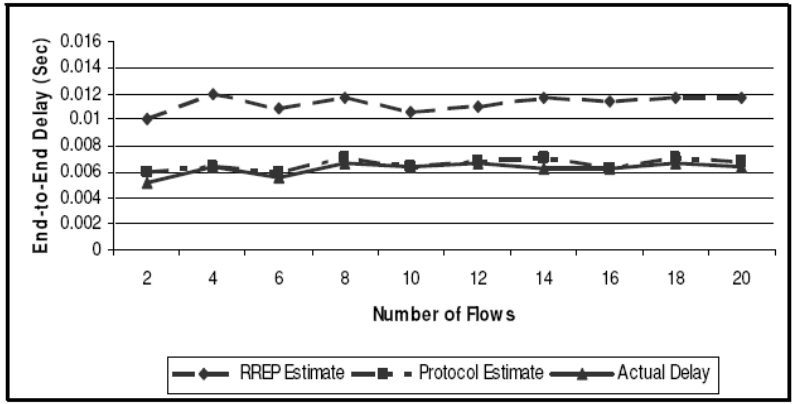}
\caption {End-to-end delay estimation by different protocols for different number of nodes in the network}
\label{Figure9}
\end{figure} 

Fig.~\ref{Figure9} shows that the probe packet-based mechanism very accurately estimates the actual delay. The naive RREP approach is very poor in estimation of the delay as explained in Section~\ref{Section4.3}.

\begin{figure}[htb]
%\vspace{2.5in}
\centering
\includegraphics [width=25pc]{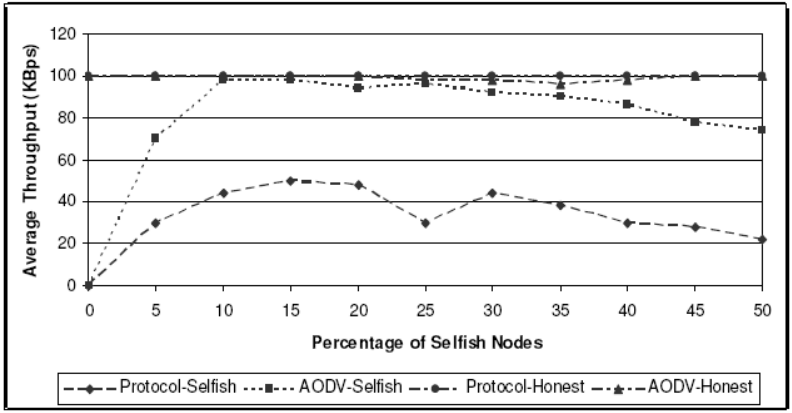}
\caption {Average throughput for different protocols with varying number of selfish nodes}
\label{Figure10}
\end{figure} 

\subsection{Detection of Selfish Nodes}\label{Section5.4}
As mentioned in Section~\ref{Section4.7}, the proposed protocol has the ability to detect selfish nodes which try to use network resources without contributing to the cooperative framework of routing. To evaluate its detection capability of selfish nodes, two types of flows are distinguished: \textit{selfish} and \textit{honest}. A flow is considered selfish if either its source or destination is a selfish node, otherwise the flow is considered to be honest. Some mesh clients are selected randomly and configured as selfish nodes. In each of the 20 run of the simulation, 10 flows of 50 kbps data rate are generated randomly. The throughput in the network is measured under four different types of flows: (i) honest flows using proposed protocol, (ii) selfish flows (flows where selfish nodes are involved) using the proposed protocol, (iii) honest flows using simple AODV protocol \cite{Perkins}, and (iv) selfish flows using AODV protocol.

It may be observed from Fig.~\ref{Figure10} that AODV cannot restrict the traffic along the selfish flows. The selfish nodes can fully exploit the routing process to have their packets routed in the network. However, the proposed protocol reduces the flows along the selfish paths. In fact, the performance of AODV is not very much affected by presence of selfish nodes, since it never establishes routing path based on hello packets. Since the proposed protocol establishes route based on hello packets received form neighbors, its performance is affected by the presence of selfish nodes. However, its performance is not substantially affected, since most in most cases, these nodes are not allowed to participate in the routing, because of the low values of their link reliability. It may also be mentioned that the proposed protocol is able to maintain, on average, 35\% more throughput in presence of selfish nodes when compared with the protocol proposed in \cite{Kone}. The large difference is due to its ability to detect selfish nodes faster by its effective bandwidth estimation in the route where non-forwarding of packets is treated as packet drops due to congestion. 

\section{Conclusion and Future Work}\label{Section6}
WMNs have become an important focus area of research in recent years owing to their great promise in realizing numerous next-generation wireless services. Driven by the demand for rich and high-speed content access, recent research has focused on developing high performance communication protocols, while security and privacy issues have received relatively little attention. However, given the wireless and multi-hop nature of communication, WMNs are subject to a wide range of security and privacy threats. Accordingly, designing a high-performance, efficient, secure and user privacy-preserving routing protocol for WMN is a very challenging task due to involvement of a number of complex factors. This paper has presented a routing protocol that has very low control overhead and high network throughput when the number of source nodes in a WMN is large. By robust estimation wireless link quality and the available bandwidth in the wireless route and exploiting the benefits of using MPRs and circular routing technique, the protocol is able to sustain a high level of throughput with a low control overhead. The user privacy is protected by using a novel anonymized authentication protocol. Simulation results have shown the protocol is more efficient than some of the existing routing protocols for WMNs. Future work includes developing a security module with the routing protocol that will be able to defend against tunnelling attack \cite{Li}, in which two malicious nodes advertise in such a way as if they have a very reliable ink between them. This is achieved by tunnelling AODV messages between the nodes. No security scheme exists so far that can detect this attack promptly and efficiently.

\end{document}